\pdfoutput=1

\documentclass[11pt, a4paper]{article}

\usepackage{xcolor}
\usepackage{framed}
\usepackage{amsmath}
\usepackage{amsfonts}
\usepackage{amssymb}
\usepackage{graphicx, rotating}
\usepackage{epsfig}
\usepackage{latexsym}
\usepackage{graphicx}
\usepackage{color}
\usepackage{amsmath,bm,amssymb}
\usepackage{cite}
\usepackage{slashed}
\usepackage{hyperref}
\usepackage{epstopdf}
\usepackage[skip=2pt]{caption} 

\usepackage{dashrule}

\usepackage{slashed}

\AppendGraphicsExtensions{.tif}
\hypersetup{colorlinks=true, citecolor=bluscuro, linkcolor=black, urlcolor=bluscuro}
\definecolor{rossos}{cmyk}{0,1,1,0.55}
\definecolor{bluscuro}{rgb}{0.15, 0.2, .85}
\definecolor{bluchiaro}{cmyk}{1,.3,0.,0.1}

\newcommand{\be}{\begin{eqnarray}}
\newcommand{\ee}{\end{eqnarray}}

\usepackage{amsmath}

\def\ba{\begin{eqnarray}}
\def\ea{\end{eqnarray}}

\def\bq{\begin{quote}}
\def\eq{\end{quote}}

\newcommand{\prr}{\prime}

\newcommand{\f}{\frac}
\newcommand{\bt}{\beta}

 at 10truept

\newcommand{\beq}{\begin{equation}}
\newcommand{\eeq}{\end{equation}}
\newcommand{\beqa}{\begin{eqnarray}}
\newcommand{\eeqa}{\end{eqnarray}}
\newcommand{\bea}{\begin{eqnarray}}
\newcommand{\eea}{\end{eqnarray}}
\newcommand{\p}{\partial}

\def\ltap{\ \raise.3ex\hbox{$<$\kern-.75em\lower1ex\hbox{$\sim$}}\ }
\def\gtap{\ \raise.3ex\hbox{$>$\kern-.75em\lower1ex\hbox{$\sim$}}\ }
\def\gl{\ \raise.5ex\hbox{$>$}\kern-.8em\lower.5ex\hbox{$<$}\ }
\def\roughly#1{\raise.3ex\hbox{$#1$\kern-.75em\lower1ex\hbox{$\sim$}}}

\def\d{{\rm d}}

\def\bt{\beta}

\newcommand{\arXiv}[2]{\href{http://arxiv.org/pdf/#1}{{\tt [#2/#1]}}}
\newcommand{\arXivold}[1]{\href{http://arxiv.org/pdf/#1}{{\tt [#1]}}}

\def\b{\bar a}

\allowdisplaybreaks

 \def\b{\beta}

\begin{document}

\renewcommand{\theequation}{\arabic{equation}}
\newcommand{\ber}{\begin{eqnarray}}
\newcommand{\eer}[1]{\label{#1}\end{eqnarray}}
\newcommand{\eqn}[1]{(\ref{#1})}
\begin{titlepage}
\begin{center}

\vspace*{2.2cm}

{\Large \bf Gauge Field Localization in the Linear Dilaton Background}

\vskip 0.6in

{\large \bf K. Farakos, A. Kehagias and G. Koutsoumbas}
\vskip 0.1in
{\em Physics Department\\
 National Technical Univeristy of Athens,  \\
     Zografou Campus, 157 80 Athens, Greece}\\
\vskip .2in

\end{center}

\vskip .6in

\centerline{\bf Abstract }

We study dynamical self-localization  of gauge theories in higher dimensions.
Specifically, we consider  a 5D  $U(1)$ gauge theory in the
linear dilaton (clockwork) background, with anisotropic gauge couplings along the transverse (fifth) direction and the longitudinal (four-dimensional) directions.  By using lattice techniques, we calculate the space plaquettes and the helicity moduli and  we
determine the phase diagram of the model. We find strong evidence that the model exhibits a
new  phase, a layer phase, where the four-dimensional  physics decouples from the
five-dimensional dynamics.   The layer phase corresponds to a strong force along the fifth direction and a Coulomb phase along the four-dimensional longitudinal directions. This is in accordance with the
 clockwork mechanism where light particles with exponentially suppressed interactions are generated in theories with no fundamental  small parameters.

\vskip 0,2cm

\vskip 4cm
\noindent
May 2020\\
\end{titlepage}
\vfill
\eject

\def\baselinestretch{1.2}
\baselineskip 16 pt
\noindent

\def\tT{{\tilde T}}
\def\tg{{\tilde g}}
\def\tL{{\tilde L}}


\section{Introduction}

The clockwork mechanism (CW)
  \cite{c1,c2,c3} provides a way to obtain light degrees of freedom with suppressed interactions in a theory that does not have small parameters.   It can be embedded in supergravity as has been shown in \cite{KR}. Various other aspects of the CW have been discussed in the  recent literature \cite{c4,c44,c5,c6,c7,c8,c9,c10,c11,c12,c13,nilles,c14,KimM,c15,c16,c17,
  c18,c19,c20,BI}.
 We will implement here, (a lattice version of) the  Continuous ClockWork (CCW) which involves an extra spacetime dimension as opposed to the Discrete Clockwork which employes a finite number of fields.   The interest in the CCW stems from the fact it is connected to the Little String Theory and moreover, it can provide a possible solution to the  naturalness problem. In particular,  dynamics of the CCW is described by the  same action as the   linear dilaton duals of Little String Theory \cite{l1,l2} with two boundary branes, in very much the same way the Randall-Sundrum (RS)  background is anti-de Sitter space time with also two boundary branes \cite{RS1,RS2}. These boundary (end of the world) branes  are located  at the fixed points of  a compactified extra fifth dimension
  $S_1/{\mathbb Z}_2$.  The minimal spectrum of the CCW theory contains a scalar field (dilaton) coupled to gravity with action  in the Einstein frame
\begin{eqnarray}
S&=&\int \d ^4x \d y\,\sqrt{-g}\left\{\frac{M_5^3}{2}
\left(R-\frac{1}{3}\partial_M S\partial^M S+4k^2 e^{-\frac{2}{3}S}\right)\right.\nonumber\\
&-&\left.\frac{e^{-\frac{1}{3}S}}{\sqrt{g_{55}}}
\left[\delta(y)\Lambda_0+\delta(y-\pi R)\Lambda_\pi\right]\right\}. \label{ccw}
\end{eqnarray}
$M_5$ is the 5D Planck mass, $R$ is  the radius extra dimension,  $\Lambda_0$ and $\Lambda_\pi$ are the brane tensions which satisfy
$\Lambda_0=-\Lambda_\pi=-4 k M_5^3$, $k^2$ is a dimensionful parameter and $0=y_0\leq y\leq y_\pi=\pi R$. The resulting  metric and dilaton is \cite{AG,c3}

\beq
\d s^2=e^{\frac{4}{3}k|y|}\left(\eta_{mn}\d x^m\d x^n+\d y^2\right), ~~~~~S=2 k |y|, \label{metccw}
\eeq
with $\eta_{mn}$ the flat Minkowski metric ($m,n=0,\cdots, 3$). Similarly with  the Randall-Sundrum (RS) case \cite{RS1,RS2}, exponential suppressions of the form  $e^{-k\pi R}$ is generated on the  $y=\pi R$ brane which leads to corresponding hierarchies.
On the other end  $y=y_0=\rm{const.}$ the CCW metric (\ref{metccw}) describes flat Minkowski spacetimes rescaled though, as in RS,  with  the factor
$e^{\frac{4}{3}k|y_0|}$.
However, contrary to the RS case where the generated hierarchy is exponential and therefore strong, here, as it has been shown in \cite{KR}, the hierarchy is only power-law \cite{Keh}. The reason is that, although there exist exponential factors entering the 4D Planck mass, the compactification radius of the internal  $S_1/{\mathbb Z}_2$ space also has an exponential dependence on $R$, contrary to the RS case where the dependence is linear.   Therefore,
the 4D Planck mass has a mild power-law dependence on the compactification radius leading to a power-law hierarchy.

It is known that massless fields can be localised on domain walls and on solitons in general. For example, vortex scalar fields may form  domain walls on which massless scalars as well as chiral fermions are localised \cite{L}--\cite{J}.  In particular, five dimensional fermions coupled to
 the vortex field, deposit a single  fermionic
zero mode on the 4D domain wall \cite{CH}.

Localization of gauge field, contrary to scalars and fermions which can easily be localised on domain walls, is quite tricky and not fully satisfactory.  The reason is simple: The charged vortex field that forms the domain wall also develops a vev and therefore breaks the 5D $U(1)$ theory. Therefore, the 5D gauge field will be massive, except possibly at the position of the domain wall where possibly the vev of the vortex field vanishes. So, we may have a massless photon localised at the wall, which however will be no capable of producing long range electric field along the wall as a result of the Meissner effect  which will give confined
magnetic flux.
Localised electric fields can be produced by reversing the situation  above
\cite{BK},\cite{DS},\cite{TR}. This amounts to have
confining medium with monopole condensation in the bulk \cite{giacomo}, which interchanges the role of the electric and magnetic fields above and confines now the electric fields with exponentially dying-off magnetic fields along the wall.

Gauge field localization can be achieved on a background with non-trivial geometry, where the role of the vortex fields is played by the geometry and the domain wall by the boundary branes. For the RS case for example, it has been shown   in \cite{KD} and \cite{dfv} by using lattice techniques, that gauge field localization is possible on the RS background due to the development of anisotropic couplings, i.e. different coupling in the transverse dimension as compared to the four space time ones. We found that in this case there exists a new phase for the 5D \cite{NF,Altes}. This new phase is the   layered phase, and corresponds to confining force in the fifth
direction and Coulomb phase in the (4D)  layers.  Therefore, both perturbatively \cite{Rizzo,GP} and non-perturbatively \cite{KD} gauge fields can be
localised on the boundary brane in RS backgrounds.
The aim of the present paper is to see whether this is also the case for the CCW. We claim that localization of gauge fields on the boundary branes is equivalent to the existence of the layer phase. Hence, we will  consider a $U(1)$ gauge theory  in the 5D linear dilaton background with the boundary branes and find its phase diagram.
Possible layer phase will show localization of the gauge degrees of freedom on the brane.

\section{$U(1)$ Lattice Gauge Fields on the CCW}

Let us consider an abelian gauge theory on the 5D  CCW background, the dynamics of which is described by the action
\begin{eqnarray}
\mathcal{S}&=&-{1\over
4 g_5^2}\int d^5 x \sqrt{-g}\, e^{ \alpha S}\, F_{MN} F_{KQ}g^{MK}g^{NQ}, \label{actn}
\end{eqnarray}
where a coupling of the dilaton to the gauge field, proportional to the dimensionless parameter $\alpha$ has been introduced.
By performing the  coordinate transformation $y\to r,$ where
\begin{eqnarray}
r=\frac{1}{\gamma k}\Big(-1+e^{\gamma k y}\Big),
\end{eqnarray}
\begin{eqnarray}
\gamma=\frac{2}{3}+2\alpha, \label{gamma}
\end{eqnarray}
the CCW metric and dilaton turns out to be
\begin{eqnarray}
ds^2=\big(1+\gamma k |r|\big)^{\frac{4}{3\gamma}}
\Big(\eta_{mn}dx^m dx^n\Big)+\big(1+\gamma k |r|\big)^{-\frac{4\alpha}{\gamma}}dr^2, ~~~S=\frac{2}{\gamma}\ln\Big(1+\gamma k r\Big), \label{nsol}
\end{eqnarray}
with
\begin{eqnarray}
0\leq r\leq r_c, ~~~~r_c=\frac{1}{\gamma k}\Big(-1+e^{\gamma k y_c}\Big).
\end{eqnarray}
Using the background solution (\ref{nsol}), the action (\ref{actn}) is written as
\begin{eqnarray}
\mathcal{S}&=&\int d^4x \int_0^{r_c}dr \,
\left(-{1\over 4 g_5^2}F_{\mu\nu}F_{\kappa\lambda}\eta^{\mu\kappa}\eta^{\nu\lambda}-{1\over
2g_5^2}\big(1+\gamma k r\big)^2 F_{\mu 5}F_{\nu 5}\eta^{\mu\nu}\right)\, .
\label{gauge0}
\end{eqnarray}

We can analytically continue the Minkowski-space action
(\ref{gauge0}) to Euclidean space so that we get
\ba
&&S_{gauge}^{E}=\int d^5x
\left({1\over
4g_5^2}F_{\mu\nu}F_{\mu\nu}+{1\over
2g_5^2}\big(1+\gamma k x_T\big)^2F_{\mu T}F_{\mu T}\right)\,
,~~~\mu,~\nu=1,...,4\, ,
\label{gauge1}
\ea
where we  have defined the original coordinate $r$ with $0\leq r\leq r_c$ as $r=x_T$
 (T denotes the fifth-transverse direction) with $0\leq x_T\leq x_{T_c}$.
Therefore, we have written this way a
gauge theory on a CCW background. We believe that all fundamental features of a a gauge theory on the CCW background is encoded in the action (\ref{gauge1}). Our aim is to study this action at the non-perturbative level and find its phases.
An essential feature is its localization properties, which we expect to be answered by studying its non-perturbative dynamics.  Clearly, the effective coupling in the fifth dimension is
\begin{eqnarray}
g_T=\frac{g_5}{1+\gamma k x_T}, \label{gt}
\end{eqnarray}
so that its value depends on the fifth dimension. In particular, we see that for
\begin{eqnarray}
-1<\gamma kx_{T_c}<0,
\end{eqnarray}
the coupling $g_T$ is growing as we go deeper in the extra dimension compared to the coupling   on  the boundary brane  at $x_T=0$. On the other hand, if
\begin{eqnarray}
\gamma kx_{T_c}>0,
\end{eqnarray}
then $g_T$ is decreasing for increasing $x_T$ and therefore, $g_T$ is growing for $x_T\to 0$ compared to the coupling   on  the boundary brane  at $x_T=x_{T_c}$.

\section{Lattice formulation}
At this point we would like to recall some facts about the construction of lattice gauge field theories.
The fundamental variables are the link variables
\be
U_M(x)=\{U_\mu=e^{ia_s {\overline A}_\mu},~U_T(x)=e^{ia_T {\overline A}_T} \},
\ee
as well as  the plaquette variables, which are defined as
\be
U_{\mu\nu}(x)=U_\mu(x)U_\nu(x+a_s {\hat \mu})
{U_\mu}^\dagger(x+a_s {\hat \nu})
{U_\nu}^\dagger(x),
\ee
\begin{eqnarray}
U_{\mu T}(x)=U_\mu(x)U_T(x+a_s{\hat \mu}){U_\mu}^\dagger(x+a_T {\hat T})
{U_T}^\dagger(x),
\end{eqnarray}
where ${\overline A}_\mu$ and
${\overline A}_T$ are the continuum gauge fields in the 4D and the transverse directions respectively and  we have denoted as $a_s,a_T$  the corresponding lattice spacings. The  dynamics is determined by the  5D standard  lattice  action
 \be
S_{\rm standard}=\b \sum_{x,1\leq \mu<\nu\leq 5}(1-Re\ U_{\mu\nu}(x,y)) .
\label{latgen}
\ee
Eq.(\ref{latgen})  assumes silently an equivalence (lattice homogeneity) between all five directions and therefore there is a single coupling $\beta$ that appears in (\ref{latgen}).
However, in our case, we have broken homogeneity as the continuous
geometry is not homogeneous and the fifth direction is different than
the remaining four. In other words, the continuous Euclidean $SO(5)$ has been broken by the underlying geometry and the  dilaton to just $SO(4)$. Therefore, we expect to have different coupling $\beta$ and $\beta_T$ along the 4D space and the transverse fith direction, respectively.  The correct action then to describe
the lattice gauge field dynamics of a (pure) gauge theory on a cubic five-dimensional lattice of  the particular non-trivial background we are discussing here,  should be of the form
\be
S=\b \sum_{x,1\leq \mu<\nu\leq 4}(1-Re\ U_{\mu\nu}(x,y))+
\sum_{x,1\leq \mu\leq 4 } \beta_T (1-Re\ U_{\mu T}(x,y))\, .
\label{lat1}
\ee
Note that  the $\beta_T$ coupling is expected to have a dependence  on the fifth coordinate $x_T=n_T a_T,$ according to the CCW scheme so that
the relations
\begin{eqnarray}
&&g_T=\frac{g_5}{1+\gamma k x_T}, ~~~~x_T=|n_T| a_T,\nonumber \\
&&  k=\f{1}{a_T},~~~ \beta_T\propto\f{1}{g_T^2},~~~  \beta\propto\f{1}{g_5^2}\label{gt1}
\end{eqnarray}
yield: \be \f{1}{g_5}=\f{1}{g_T}\frac{1}{1+\gamma k x_T} \Rightarrow  \beta = \f{\beta_T}{(1+\gamma |n_T|)^2}.\label{bbpr}\ee
We remark that for $\b\neq \beta_T$ there exists just a 4D Poincar\'e invariance in the continuum limit which however is enhanced to 5D Poincar\'e for $\b=\beta_T.$

The na\"ive continuum limit of the theory is obtained in the limit   $a_s\to 0$ and $a_T\to 0$. In this limit, the lattice action (\ref{lat1}) degenerates to
\be
S_{\rm c}= \sum_{x,1\leq \mu<\nu\leq 4} {\bt \over 2} F_{\mu\nu}^2
+ \sum_{ x,1\leq \mu\leq 4 } \f{\beta_T}{2} F_{\mu T}^2 + {\cal
O}(a^5),
\ee
with
\begin{eqnarray}
&&
F_{\mu \nu}(x) \equiv A_\nu(x+a_s {\hat \mu})-A_\nu(x)
-A_\mu(x+a_s {\hat \nu})+A_\mu(x),\nonumber \\
&&F_{\mu T}(x) \equiv A_T(x+a_s {\hat \mu})-A_T(x)
-A_\mu(x+a_T {\hat T})+A_\mu(x).
\end{eqnarray}
Next we define the (continuum) fields $\overline{A}_\mu$ and and
$\overline{A}_T$ (we will generally denote field in the continuum with a bar):
$$
~~ {\overline A_\mu}=\frac{A_\mu}{  a_s}, ~~ ~(\mu=1,2,3,4),~~~~~~{\overline A_T}=\frac{A_T}{ a_T}.
$$
The mixed transverse-4D part of the  gauge action may then  be
rewritten in the form:
\begin{eqnarray}
&&\sum_x \f{\beta_T}{2} a_T^2 a_s^2 \sum_{\mu=1,2,3,4} ({\overline F}_{\mu T})^{2}=\nonumber \\
 &&\hspace{3mm} =\sum_x \f{\beta_T a_{T}}{2 a_s^2}  a_{s}^{4}a_{T} \sum_{\mu=1,2,3,4} ({\overline F}_{\mu T})^{2}
\rightarrow  \int d^{5}x \f{\beta_T a_T}{2 a_s^2}
\sum_{\mu=1,2,3,4} ({\overline F}_{\mu T})^{2},
\end{eqnarray}
whereas the pure 4D part is
\begin{eqnarray}
\sum_x \f{\beta}{2 a_T} a_s^4 a_T
\sum_{1 \le \mu < \nu \le 4} ({\overline F_{\mu \nu}})^2
\rightarrow \sum_{1 \le \mu < \nu \le 4}\int d^5 x
\frac{\beta}{2 a_T} ({\overline F_{\mu \nu}})^2.
\end{eqnarray}
From this point on we specialize to $a_s=a_T\equiv a.$
The action in the continuum limit can be written as
\begin{eqnarray}
S_{\rm c}= \int d^5 x {\left [ \f{1}{4 g_{5}^{2}}
\sum_{1 \le \mu < \nu \le 4} ({\overline F_{\mu \nu}})^{2}
+\f{1}{2 g_{T}^{2}}\sum_{\mu=1,2,3,4} ({\overline F_{\mu T}})^{2} \right ]},
\end{eqnarray}
where $\b$ and $\b_T$ are given explicitly by
\be
\beta \equiv \frac{a}{g_{5}^{2}},
~~\beta_T \equiv \frac{a}{g_{T}^{2}}.
\ee
The overall coupling $g_5$ has mass dimension $-1/2$ and it is related to the typical scale of the extra dimension.
%
%

The algorithm used  for the simulation is a  5-hit Metropolis  augmented  by an overrelaxation method. The latter amounts to  express the action for a four-dimensional contribution as $\beta C cos(\phi+\theta_\mu)$
and writing
\begin{eqnarray}
 C \cos \phi & =& \sum \cos \chi_s + q \sum \cos \chi_T, \nonumber \\
 C \sin \phi & =&  \sum \sin \chi_s + q \sum \sin \chi_T,
\end{eqnarray}
where we denote the space-like staples with $\chi_s,$ while the transverse-like ones as  $\chi_T.$ In addition,  $q \equiv \f{\bt_T}{\bt},$ is the ratio of the two couplings $\beta$ and $\beta_T$ when
$\theta_\mu$ is longitudinal (four-dimensional)  and 1 otherwise. Then, the overrelaxation method \cite{creutz,brown} determines also the new link which turns out to be $\theta'_\mu = -\theta_\mu-2 \phi.$ For transverse-like $\theta_T,$ only $\chi_T$ come into play and $\theta_T \to -\theta_T-2 \phi$ again. The change is always accepted.


\section{Results}

%
%

\subsection{Observables}

Let us describe the quantities we use to spot the phase transitions.
An operator that we use heavily in this work is the mean value $\langle
\hat{P}_s\rangle$ of the space-like plaquette, defined through:
\be
{\hat P}_s(|n_T|) \equiv \f{1}{12 N^4}
\sum_{x,1 \le \mu < \nu \le 4} \left.\cos F_{\mu \nu}(x)\right|_{{\rm fixed} |n_T|}
\ee
In addition we consider the helicity modulus, used in \cite{VF}. This quantity is expected to vanish in the strong coupling phase  and assume non-zero values in the Coulomb phase (\cite{JLC}).
It is defined via \be h(\beta)=\left.\f{\p^2 F(\Phi)}{\p \Phi^2}\right|_{\Phi=0},\ee where $\Phi$ is the flux of an external magnetic field and $F(\Phi)$ is the corresponding free energy.
The definitions read: \be F(\Phi)=-\ln[Z(\Phi)],\ \ Z(\Phi)=\int D\theta \exp\left[\sum_{S} \beta \cos(\theta_P+\Phi)+\sum_{\overline{S}} \beta \cos(\theta_P)\right],\ee where the sum extends over a set $S$ of plaquettes with a given orientation $(\mu\nu$ for example), on which an extra flux $\Phi$ is imposed, while its complement, $\overline{S},$ consists of the plaquettes for which no additional flux is present. For example, if we choose $\mu=1,\ \nu=2,$ and fix the values for $x$ and $y,$ this set is the collection of all the plaquettes $\theta_P(x=1,y=1,z,t;\mu=1,\nu=2).$ Following the definition we find:
\be h(\beta)=\f{1}{(L_\mu L_\nu)^2} \left[\left<\sum_{\mu\nu\ plaquettes} \beta \cos(\theta_P)\right> - \left<\left(\sum_{\mu\nu\ plaquettes} \beta \sin(\theta_P)\right)^2\right>\right].\ee The symbol $\langle\ldots\rangle$ denotes the statistical average.


In this work we calculate the helicity modulus associated with the $\mu=1,\ \nu=2$ orientation, which, of course, also depends on $|n_T|.$ This quantity is defined through:
\be {\hat h}_s(\beta,\ |n_T|) \equiv \f{1}{(L_\mu L_\nu)^2} \sum_{x} \left(\Big<\sum_{P_s} (\beta \cos \theta_{\mu \nu})\Big>-\Big<\sum_{P_s} (\beta \sin \theta_{\mu \nu})^2\Big>\right),\ee where we denote by $P_s$ the plaquettes with the $(1,2)$ orientation.

\subsection{Results}

We have simulated the action (\ref{lat1}) on the lattice. A similar model with anisotropic couplings, which are constant everywhere in the lattice, has been studied in \cite{KD} and \cite{dfv}.
 In particular,  the phase diagram is given below in Fig.1.
%

\begin{center}
\begin{figure}[!h]
\centering
\includegraphics[scale=0.6]{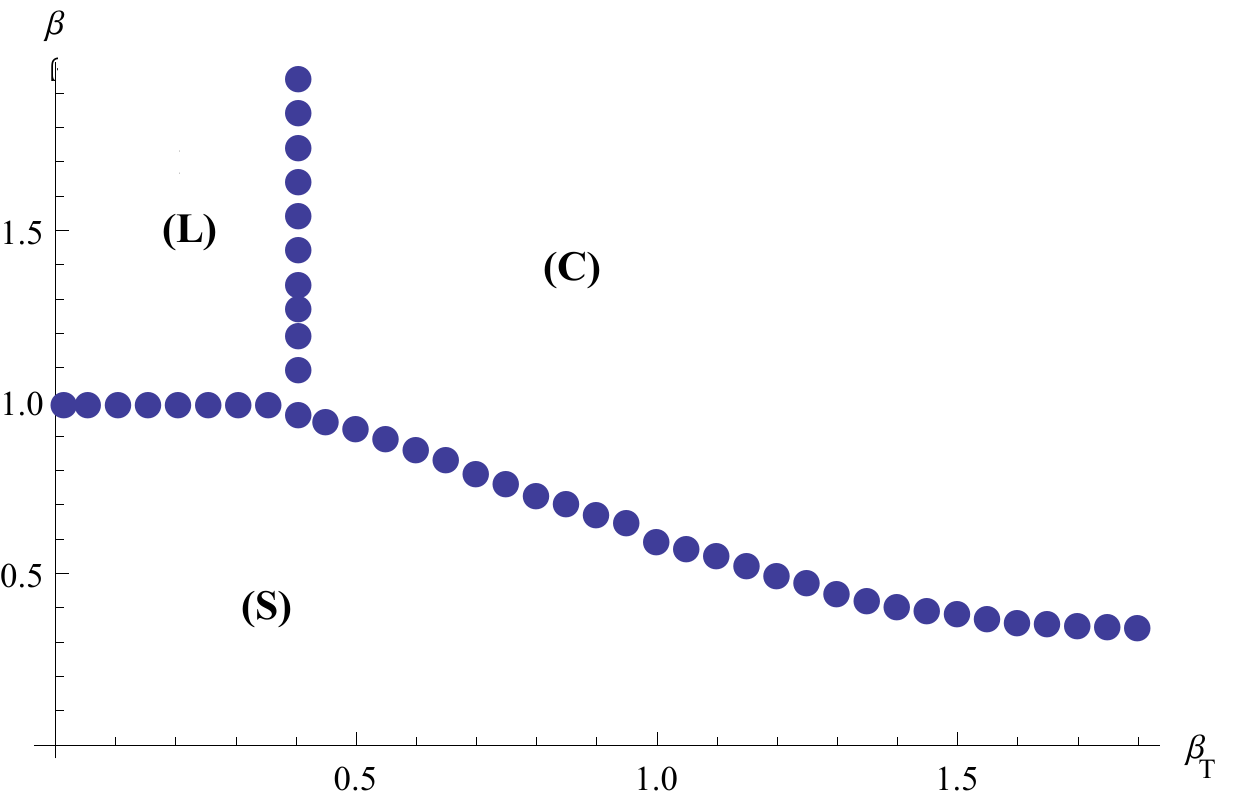}
\caption{The phase diagram of pure, $5D$  U(1) gauge theory. The theory has constant anisotropic couplings in the longitudinal and transverse fifth direction.  The region in the lower part corresponds to the strong phase, denored by $(\mathbf{S})$. The upper left part, corresponding to small values for $\beta_T$ and relatively large values for $\beta,$ describes the layer phase, denoted by $(\mathbf{L})$. Finally, the upper right part, with large values for both $\beta$ and $\beta_T,$ corresponds to the Coulomb phase 
($\mathbf{C}$) of the gauge theory.}
\label{f1}
\end{figure}
\end{center}

We note in particular the appearance of  new phase, the so-called layer phase.
The physics of the latter can be understood as follows. Let us consider a five-dimensional gauge theory in the  Coulomb  phase where both $\beta$ and $\beta_T$ are large; charged particles in the  five dimensional ambient space experience a Coulomb force.   Next, we keep $\beta$ constant and decrease the value of $\beta_T.$ Since $\beta$ is kept fixed, there will still be a Coulomb force along the four longitudinal directions. However, as the coupling $\beta_T$ is decreasing, there will be a critical value for $\beta_T,$ where the force along the fifth direction will be strong enough to allow for a confining force; the force along the four longitudinal directions is always of a Coulomb nature. This gives rise to the layer phase. If, keeping $\beta_T$ small enough, one also decreases $\beta,$ there will appear a strong confining force also along the four longitudinal directions as well, when $\beta$ gets smaller than some critical value. Thus we reach a strong coupling phase, where confining forces act along all directions. The above will be manifest in the Wilson loops, which are ultimately connected to the potential between test particles. Therefore, the expected behaviour of the Wilson lines are as follows:
\begin{eqnarray}
\begin{tabular} {lll}
\mbox{1.~Strong phase} ($\mathbf{S}$)&$\to$&$W_{\mu \nu}(L_1,L_2) \approx \exp(-\sigma L_1 L_2) $,
\\
&&\\
\mbox{2.~Coulomb phase}  ($\mathbf{C}$)&$\to$&$W_{\mu \nu}(L_1,L_2) \approx \exp(-\tau (L_1+L_2))$,
\\
&&\\
\mbox{3.~Layer phase} ($\mathbf{L}$)&$\to$& $
\left\{\begin{tabular}{l}
$W_{\mu \nu}(L_1,L_2) \approx \exp(-\tau (L_1+L_2))$,\\
$W_{\mu T}(L_1,L_2) \approx \exp(-\sigma^\prr L_1 L_2)$,
\end{tabular}\right.
$
\end{tabular}
\end{eqnarray}
where $\sigma,~\tau,~\sigma^\prr$ are dimensionful (positive) constants.
Clearly, the layer phase is due to the fact that the theory  can be in different phases in the transverse and longitudinal directions. Namely, the layer phase is manifestation of the theory being   confining in the fifth direction while being Coulombic in the rest. Therefore, a layer phase exists in a theory that exhibits both strong and Coulomb phase and therefore a  non-Abelian gauge theory may display a layer phase in  six dimensions at least.

Let us note that there is no layer phase for a gauge theory realized  by a 4D Coulomb phase  in the longitudinal directions and a Higgs phase  in the transverse one through an appropriate Higgs mechanism. The reason is that  there cannot be a Coulomb phase along the logitudinal 4D directions   due to the Meissner effect which demands   an exponential die off of the 4D electric fields,  and therefore leads to the lack of massless photon in the longitudinal directions. Note that there exist higgs models, \cite{SFDKK}, with a layer phase, that is Coulomb or Higgs phases in the 4D space along with a strong coupling phase along the transverse direction. Non-abelian examples may be found in \cite{DFK} and \cite{DS}.

We have chosen  to probe the phase transition $(\mathbf{S} \leftrightarrow \mathbf{L})$ between  the strong and the layered phase and the transition $(\mathbf{S} \leftrightarrow \mathbf{C})$ between the strong and the Coulomb phase.
To this end, we fixed the transverse coupling to the value $\beta_T=0.2$ for the $\mathbf{S} \leftrightarrow \mathbf{L}$ transition and to the value $\beta_T=1.2$ for the $\mathbf{S} \leftrightarrow \mathbf{C}$ transition, then we varied the space-time coupling $\beta$ so that equation (\ref{bbpr}) is satisfied.

The columns of the table in equation (\ref{mat0212}) contain the values of $\beta$ for various values of $\gamma.$ In the first column we give the number of the hyperplane coordinate $n_T$ in the transverse direction. Since we work with a $10^4\times 16$ lattice, we number the sites from $0$ to $8$ and notice that sites $n_T=9, \dots, 15$ may also be represented by the differences $n_T-16,$ whose absolute value is the distance of the relevant site from the site at $n_T=0.$ Thus we consider a site of reference at $n_T=0,$ which coincides with $n_T=16,$ two sites, at $n_T=1$ and at $n_T=15\rightarrow 15-16=-1$ at distance $|n_T|=1,$  two sites, at $n_T=2$ and at $n_T=14\rightarrow 14-16=-2$ at distance $|n_T|=2,$ up to the sites at $n_T=7$ and at $n_T=9\rightarrow 9-16=-7$ at distance $|n_T|=7.$ The site at $n_T=8$ lies at the largest distance, $|n_T|=8,$ from the reference site.

The second column contains the values of $\beta$ obtained for $\gamma=-0.075$ and $\beta_T=0.2.$ One may observe that the four-dimensional volumes at each $n_T$ have couplings $\beta$ start with $\beta= \beta_T$ at the reference site $|n_T|=0$ and they get bigger and bigger values for $\beta$ for larger distances $|n_T|;$ the largest value is achieved for $|n_T|=8.$ Thus the system is expected to lie initially in the strong phase and move towards the layered phase for larger distances. This behaviour is due to the negative value of $\gamma$ and is repeated, with quantitative changes for $\gamma=-0.085$ and $\gamma=-0.095.$

For $\gamma=0.150$ and $\gamma=0.250$ the behaviour is different: the system is expected to start off in the Coulomb phase at small distances and move towards the strong coupling phase for larger distances, where the values of $\beta$ become small.

\begin{eqnarray}
\begin{tabular}{||c|c|c|c|c|c||}
\hline
 \multicolumn{6}{|c|}{Coupling $\beta$ for various values of $\gamma$ and
 $n.$} \\
 \hline
\hline\hline n & $\gamma=-0.075$ & $\gamma=-0.085$ & $\gamma=-0.095$ & $\gamma=0.150$ & $\gamma=0.250$
\\
\hline
\hline
$0$ & $0.20$ & $0.20$ & $0.20$ & $1.20$ & $1.20$
\\ $1$ & $0.23$ & $0.24$ & $0.25$ & $0.91$ & $0.77$
\\ $2$ & $0.28$ & $0.29$ & $0.30$ & $0.71$ & $0.53$
\\ $3$ & $0.33$ & $0.36$ & $0.39$ & $0.57$ & $0.39$
\\ $4$ & $0.41$ & $0.46$ & $0.52$ & $0.47$ & $0.30$
\\ $5$ & $0.51$ & $0.60$ & $0.73$ & $0.39$ & $0.24$
\\ $6$ & $0.66$ & $0.83$ & $1.08$ & $0.33$ & $0.19$
\\ $7$ & $0.89$ & $1.22$ & $1.78$ & $0.29$ & $0.16$
\\ $8$ & $1.25$ & $1.95$ & $3.47$ & $0.25$ & $0.13$
\\ $9  \rightarrow -7$ & $0.89$ & $1.22$ & $1.78$ & $0.29$ & $0.16$
\\ $10 \rightarrow -6$ & $0.66$ & $0.83$ & $1.08$ & $0.33$ & $0.19$
\\ $11 \rightarrow -5$ & $0.51$ & $0.60$ & $0.73$ & $0.39$ & $0.24$
\\ $12 \rightarrow -4$ & $0.41$ & $0.46$ & $0.52$ & $0.47$ & $0.30$
\\ $13 \rightarrow -3$ & $0.33$ & $0.36$ & $0.39$ & $0.57$ & $0.39$
\\ $14 \rightarrow -2$ & $0.28$ & $0.29$ & $0.30$ & $0.71$ & $0.53$
\\ $15 \rightarrow -1$ & $0.23$ & $0.24$ & $0.25$ & $0.91$ & $0.77$
\\ \hline \hline
\end{tabular}  \label{mat0212}
\end{eqnarray}

\begin{center}
\begin{figure}[!h]
\centering
\includegraphics[scale=0.6]{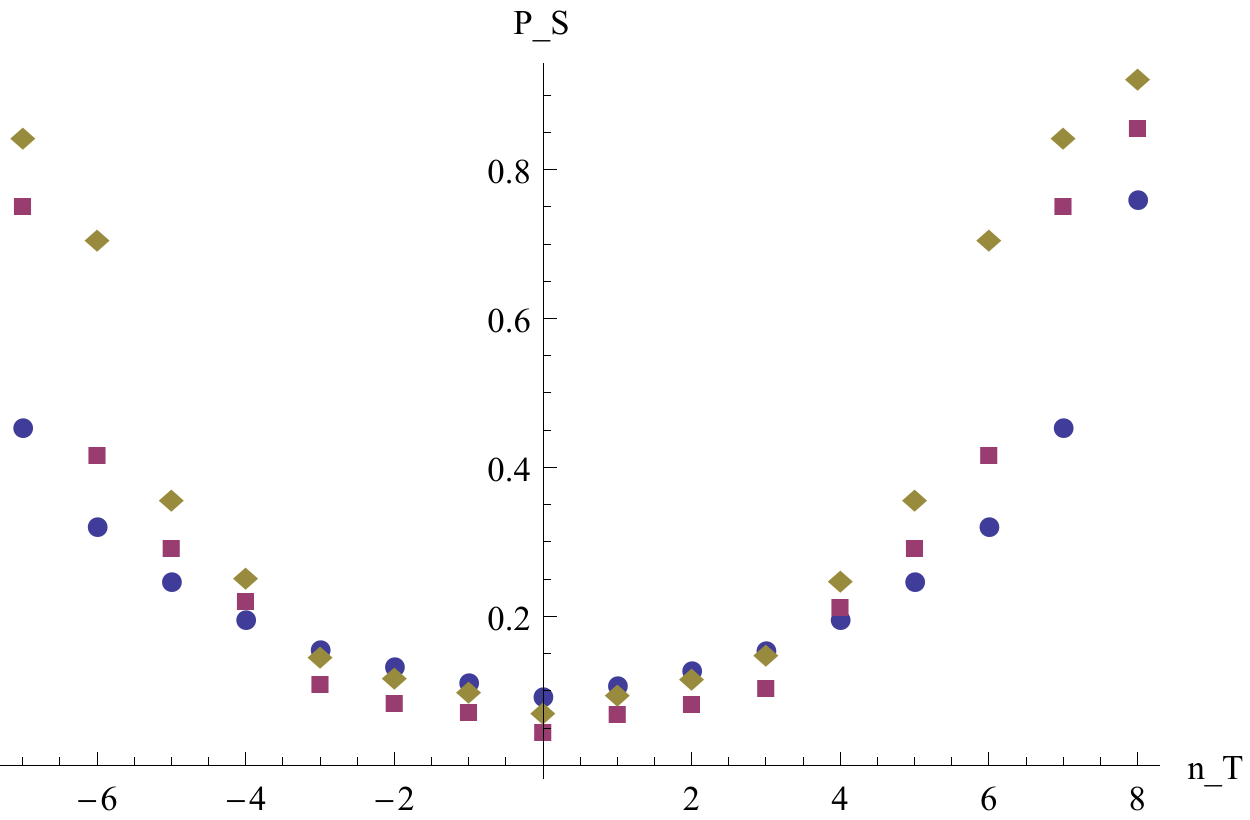}
\includegraphics[scale=0.6]{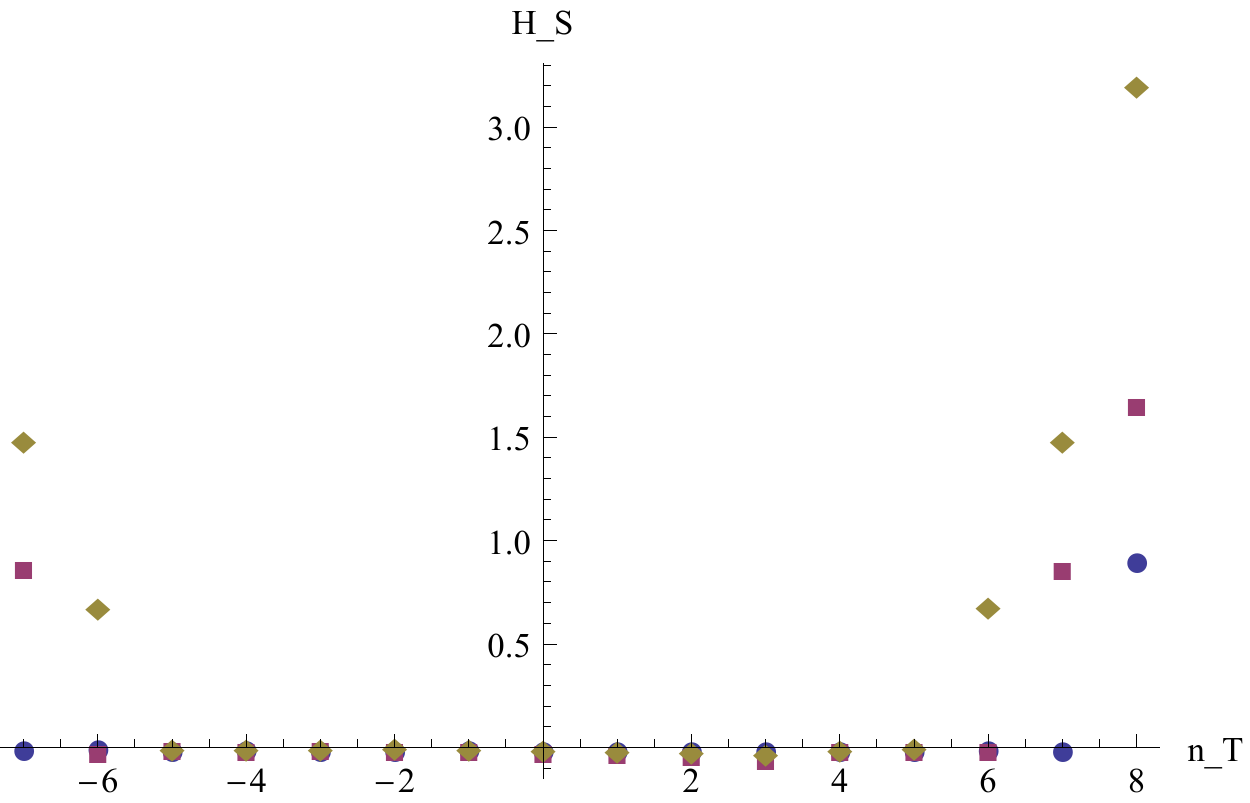}
\caption{(a) Left: Variation of space plaquettes $P_S$ versus $n$ for $\beta_T=0.2$ and various values for $\gamma.$ The lattice dimensions are $10^4\times 16.$ Blue circles correspond to  $\gamma=-0.075,$ purple squares to $\gamma=-0.085$ and brown rhombuses correspond to  $\gamma=-0.095.$ (b) Right: Variation of helicity moduli $H_S$ versus $n$ for $\beta_T=0.2$ and various values for $\gamma.$ Blue circles  correspond to  $\gamma=-0.075,$ purple squares to $\gamma=-0.085$ and brown rhombuses correspond to  $\gamma=-0.095.$}
\label{PSHS02}
\end{figure}
\end{center}

For $\beta_T=0.2$ we start with $\beta=\beta_T$ at $n_T=0$ and then we use negative values for $\gamma,$ so that $\beta$ gets big enough to cross the phase transition point, which lies at about $\beta=1.0$ for $\beta_T=0.2,$ according to the results of \cite{KD}. We show the results for the plaquettes at $\gamma=-0.075,$ $\gamma=-0.085$ and $\gamma=-0.095$ in the left panel of figure \ref{PSHS02}. We find plaquette values corresponding to the strong phase at the hyperplanes surrounding $n_T=0,$ while for large $|n_T|$ the plaquette values are consistent with a Coulomb phase.

One would like to have a more exact criterion distinguishing the phases. To this end we employ the helicity moduli, which are expected to vanish in the strong phase and  take on non-zero values in the Coulomb phase. The results are depicted in the right panel of figure \ref{PSHS02}. We observe in this figure that, for $\gamma=-0.075,$ only at $|n_T|=8,$ i.e. $\beta=1.25$ one gets  non-zero value for the helicity modulus. For $\gamma=-0.085,$ one gets non-zero values for both $|n_T|=7$ and $|n_T|=8,$ corresponding to $\beta=1.22$ and $\beta=1.95$ respectively. Finally for $\gamma=-0.095,$ one gets non-zero values for $|n_T|=6,$ $|n_T|=7$ and $|n_T|=8,$ corresponding to $\beta=1.08,$ $\beta=1.78$ and $\beta=3.47.$ We observe that all three phase changes occur at $\beta\ge 1,$ where the phase transition point is expected for the anisotropic model with constant $\beta$ and $\beta_T.$ This is exactly what one would guess for $\beta_T=0.2,$ since for this value the layers are expected to be unrelated to one another, so the fact that $\beta$ is different for each hyperplane makes little difference.

Then we will fix the transverse coupling to the value $\beta_T=1.2$ and vary the space-time coupling $\beta,$ so that equation (\ref{bbpr}) is satisfied. We start with $\beta=\beta_T$ at $n_T=0$ and then we have to use positive values for $\gamma,$ so that $\beta$ gets small and crosses the phase transition point, which lies at $\beta\approx 0.50$ for $\beta_T=1.2.$ The results are depicted in the following figures. We see in the left panel of figure \ref{HSPS12} that the plaquette takes values pertaining to the Coulomb phase in the neighbourhood of $n_T=0$, where also the $\beta$ values are large, while, at $|n_T|$ sufficiently large, the values are compatible with the strong phase. The differences in the plaquette values are not very conclusive concerning the identity of the relevant phases, so once more we will use the corresponding results for the helicity moduli, which are depicted in the right panel of figure \ref{HSPS12}. We spot non-zero values, signalling a Coulomb phase for $|n_T|\le 3,$ while the remaining sites lie in the strong coupling phase. For $\gamma=0.150$ the Coulomb phase is found for $\beta\ge 0.57,$ while for $\gamma=0.250$ the Coulomb phase is found for $\beta\ge 0.39.$ It should be noted that $\beta_T=1.2,$ so that the layers are expected to interact with one another. Thus one finds out that, although $\beta = 0.39$ at $|n_T|= 3$ describes a system in the Coulomb phase for $\gamma=0.250$, the (equal) value $\beta = 0.39$ at $|n_T|= 5$ lies deeply into the strong phase for $\gamma=0.150.$
\begin{center}
\begin{figure}[!h]
\centering
\includegraphics[scale=0.6]{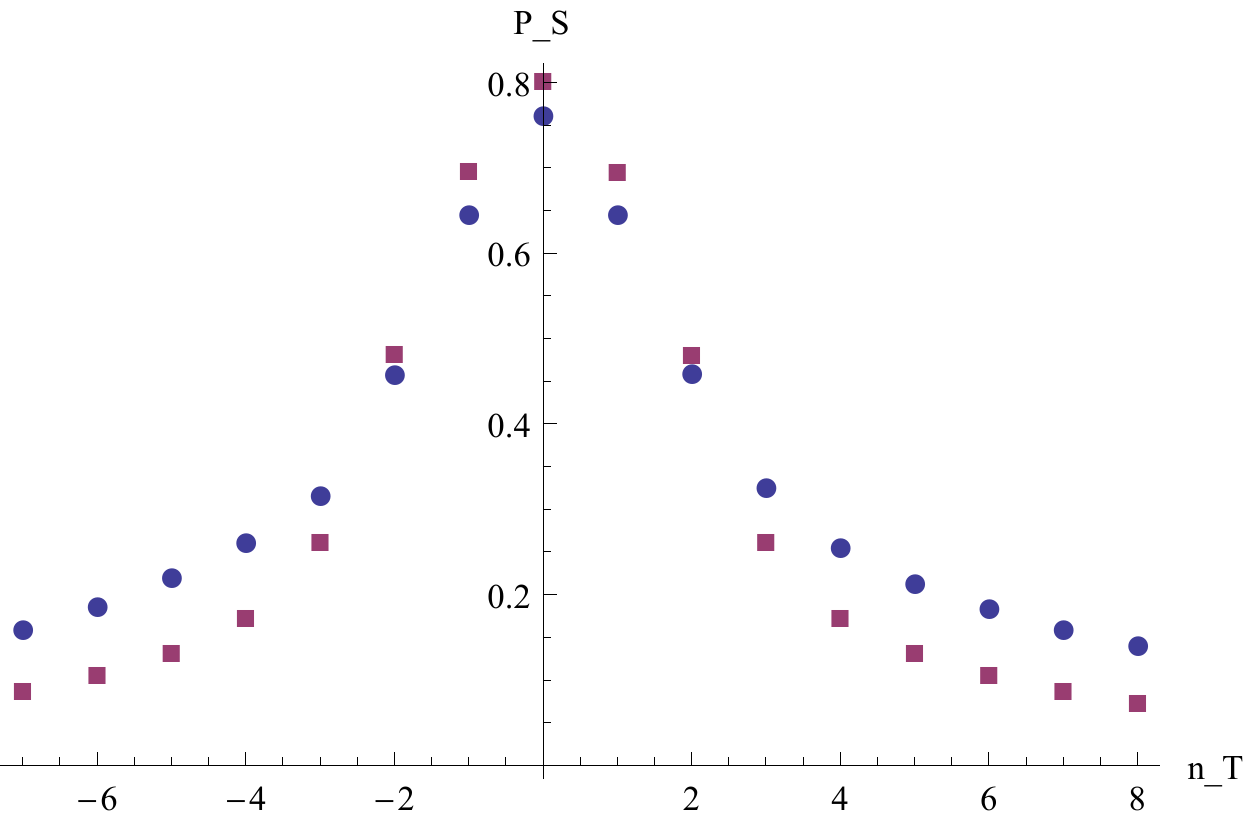}
\includegraphics[scale=0.6]{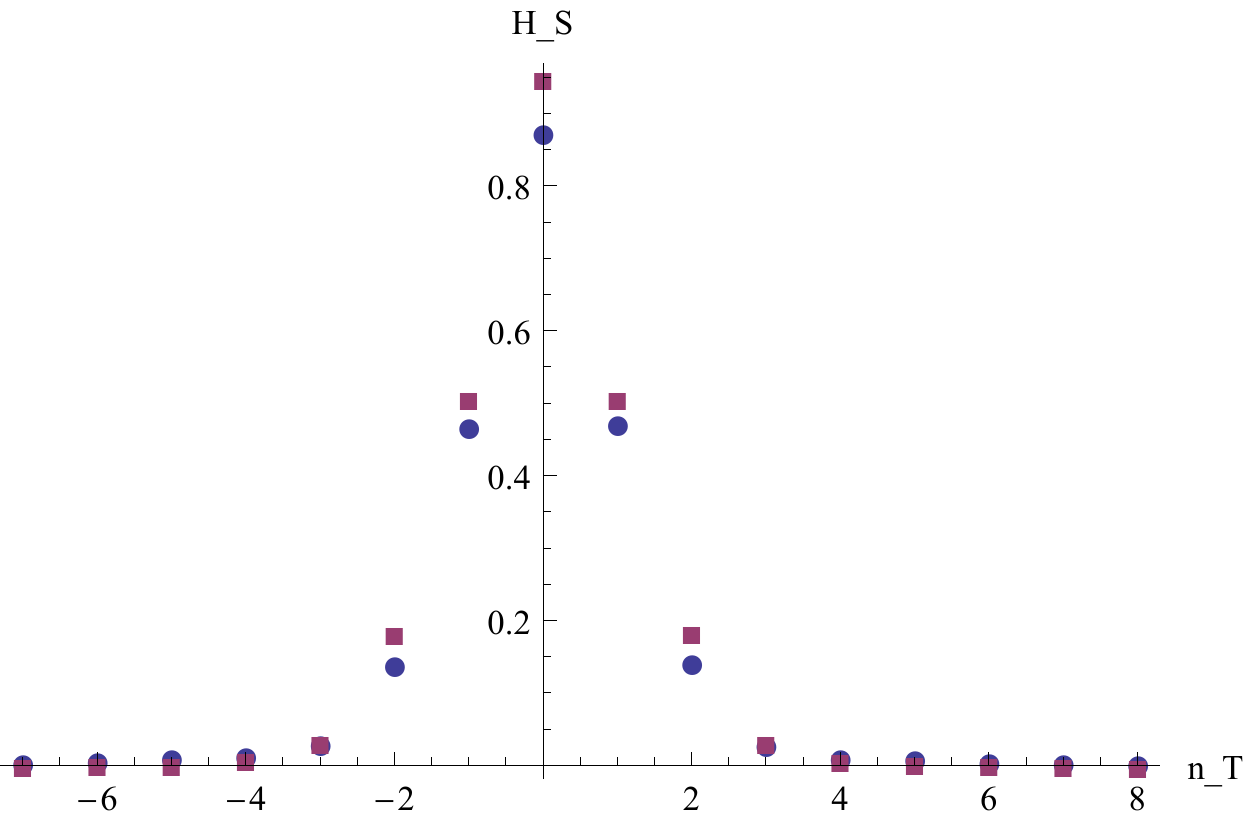}
\caption{(a) Left: Variation of space plaquettes $P_S$ versus $n_T$ for $\beta_T=1.2$ and various values for $\gamma.$ The lattice dimensions are $10^4\times 16.$ Blue circles  correspond to $\gamma=0.150, $ while the purple squares to $\gamma=0.250.$ (b) Right: Variation of helicity moduli $H_S$ versus $n_T$ for $\beta_T=1.2$ and various values for $\gamma.$ The blue circles correspond to $\gamma=0.150, $ while the purple squares to $\gamma=0.250.$}
\label{HSPS12}
\end{figure}
\end{center}
To facilitate comparison against the case with $\beta_T=1.2,$ we will depict the results of figure \ref{PSHS02} in a sightly different fashion. We start with the remark that the lattice is periodic in all directions, in particular the transverse one. Thus one may get the part of figure \ref{PSHS02} between $n_T=-7$ and $n_T=0$ and transfer it to the right of the part between $n_T=1$ and $n_T=8.$ In other words, for the left half of the graph we change $n_T$ to $16+n_T.$ In this way figure \ref{PSHS02} becomes figure \ref{PSHS02alter}.

Comparing figures \ref{HSPS12} and \ref{PSHS02alter} we observe that they are qualitatively different, since in the former the layers are highly correlated with each other, while in the latter they are independent. For instance in the right panel of the former figure we see that the values for the helicity moduli are very close to one another, despite the difference in $\gamma$'s, which corresponds to different $\beta$'s. This behaviour is quite different in figure \ref{PSHS02alter}, where different $\gamma$'s, result in a serious differences in the values for the helicity moduli. It seems that, for $\beta_T=1.2,$ there exists a correlation length in the transverse direction, which has a very mild dependence on $\gamma$. There is no correlation for $\beta_T=0.2$.

\begin{center}
\begin{figure}[!h]
\centering
\includegraphics[scale=0.6]{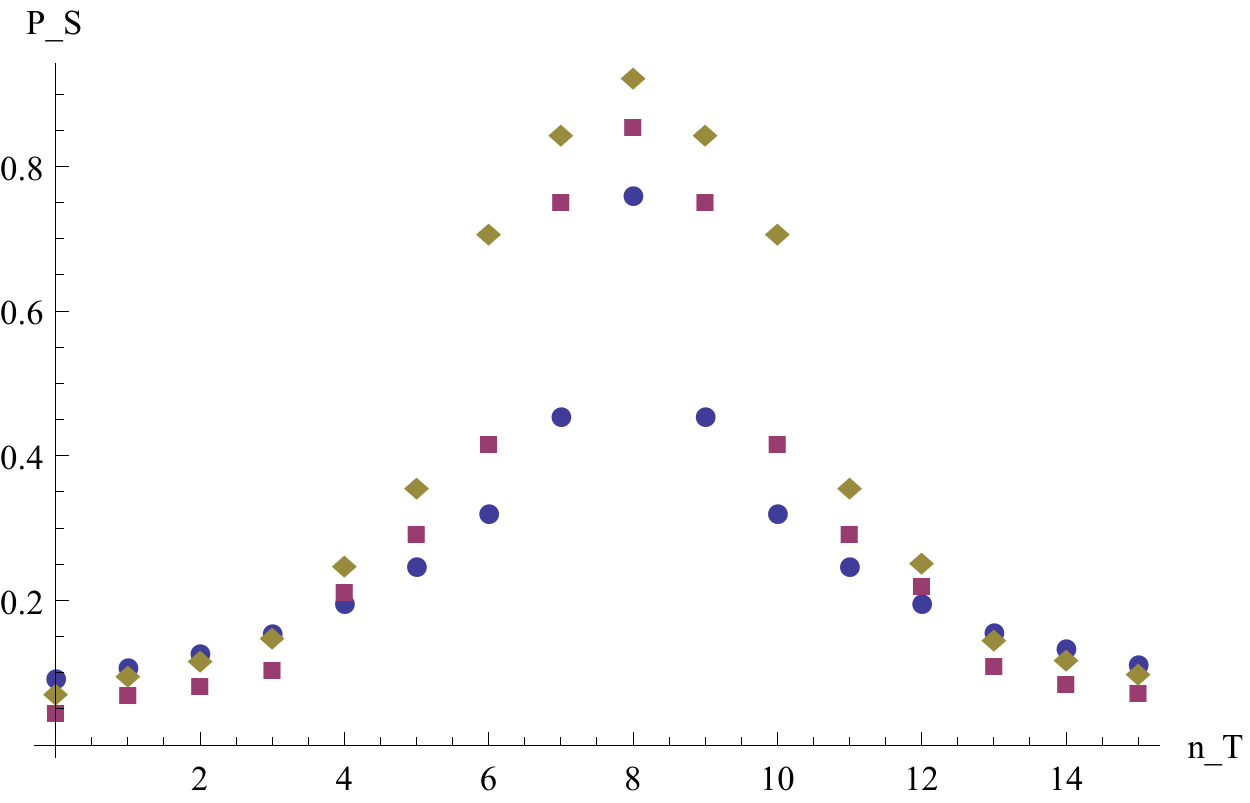}
\includegraphics[scale=0.6]{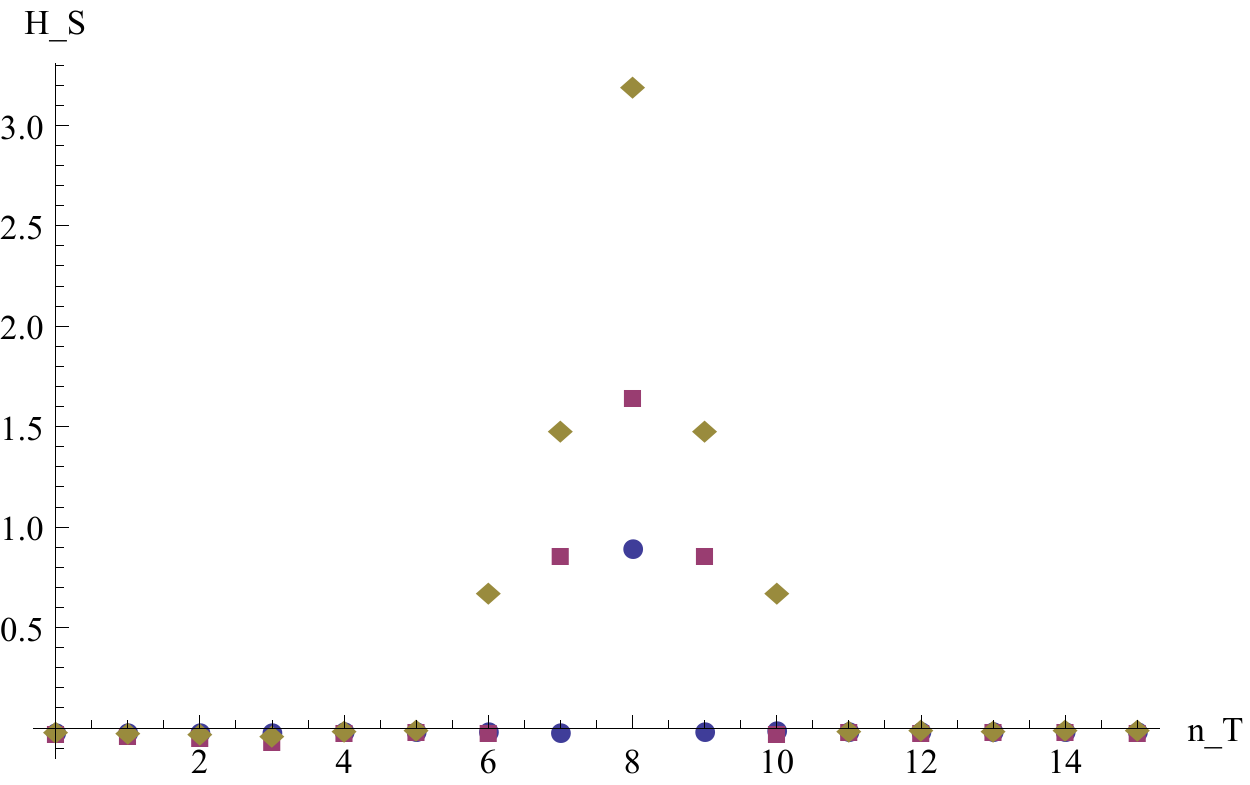}
\caption{(a) Left: Reproduction of part (a) of figure \ref{PSHS02}, corresponding to $\beta_T=0.2,$ in a  different interval of the independent variable. Blue circles correspond to  $\gamma=-0.075,$ purple squares to $\gamma=-0.085$ and brown rhombuses correspond to  $\gamma=-0.095.$ (b) Right: Reproduction of part (b) of figure \ref{PSHS02}. Blue circles  correspond to  $\gamma=-0.075,$ purple squares to $\gamma=-0.085$ and brown rhombuses correspond to  $\gamma=-0.095.$}
\label{PSHS02alter}
\end{figure}
\end{center}

As a final comment, let us determine the KK spectrum of a $U(1)$ theory in the
clockwork background.
The equation of motion for a massless photon is  in this case
\begin{eqnarray}
\partial_M\Big(\sqrt{-g}g^{MN}g^{KL}e^{\alpha S}F_{NK}\Big)=0
\end{eqnarray}
which for $A_5=0$ and in  the Lorentz gauge $\nabla_\mu A^\mu=0$
reduces to
\begin{eqnarray}
\eta^{mn}\partial_\mu\partial_\nu A_\mu+e^{-\gamma k|y|}\partial_5\Big(e^{\gamma k|y|}\partial_5\Big) A_\mu=0. \label{AA}
\end{eqnarray}
Expressing $A_\mu$ as $A_\mu(x^\mu,y)=e^{-ip_\mu x^\mu}e^{-\gamma k |y|/2} V_\mu(y)$, we find that $V_\mu(y)$ satisfies
\begin{eqnarray}
V_\mu''-\left(\frac{k^2\gamma^2}{4}+\gamma k \delta(y)-k\gamma\delta(y-y_\pi)\right)V_\mu=-m_n^2 V_\mu,
\label{psin}
\end{eqnarray}
where $p_\mu p^\mu=-m_n^2$.
Therefore,  (even) eigenvectors $V_{\mu}^n$ satisfy then the boundary conditions
\begin{eqnarray}
{V_{\mu}^n}'-\frac{\gamma k}{2}V_{\mu}^n\Big|_{y=0,y_\pi}=0.
\end{eqnarray}
In particular, the boundary condition at $y=0$ gives
\begin{eqnarray}
V_{\mu}^n=\frac{1}{N_n}\left\{\cos\Big(\sqrt{m_n^2-\frac{k^2\gamma^2}{4}}\, |y|\Big)+\frac{k\gamma}{2\sqrt{m_n^2-\frac{k^2\gamma^2}{4}}}\sin\Big(\sqrt{m_n^2-\frac{k^2\gamma^2}{4}}\, |y|\Big)\right\}, \label{eig}
\end{eqnarray}
whereas the condition at $y=y_\pi$ leads to
\begin{eqnarray}
\sin(\sqrt{m_n^2-\frac{k^2\gamma^2}{4}}\,y_\pi)=0
\end{eqnarray}
which  specifies  the KK spectrum to be
\begin{eqnarray}
m_n^2=\frac{n^2\pi^2}{y_\pi^2}+\frac{k^2\gamma^2}{4}, ~~~n=\pm1,\pm2,\ldots
\end{eqnarray}
 The same spectrum is also found for odd eigenfunctions with Dirichlet boundary conditions.

 Note that the zero mode $A_\mu^0(x,y)$, which corresponds to  $n=0$, is just
\begin{eqnarray}
 A_\mu^0(x,y)=A_\mu^0(x) \label{a0}
 \end{eqnarray}
 i.e., independent from the fifth direction. Indeed, taking the $m_n\to 0$
 limit of (\ref{eig}) we get $V_\mu\sim e^{\gamma k|y|/2}$ which leads to
 (\ref{a0}).
Note that the energy density $\rho(x,y)=-{T_0^0}$ of the zero-mode turns out to be
\begin{eqnarray}
\rho(x,y)=e^{-\frac{8k}{3}|y|} \rho(x)
\end{eqnarray}
which is localized around $y=0$.
This is in accordance with our findings for the helicity modulus, which expresses the response of the free energy to an external magnetic field.

\section{Conclusions}
We have study here the self-localization of a $U(1)$ gauge theory in a 5D background. The latter is the clockwork background which is just the 5D linear-dilaton
with two branes of different and opposite tensions at a finite distance of each other.  We allow interactions of the dilaton to the gauge field and we have seen that  the couplings in the longitudinal four-dimensions and in the fifth transverse dimension are different. In other words, the background geometry introduces anisotropic couplings and naturally splits the dynamics
into longitudinal and transverse.   This allows for non-trivial dynamics, which
leads to different phases for the gauge theory. To study the gauge dynamics, we have used lattice techniques. In particular, we have calculated the space plaquettes and the helicity moduli in order to  determine  the phase diagram of the model. We found that there is a strong phase and we provided evidence that  the model exhibits a
new  phase, the layer phase. The latter  describes pure four-dimensional  physics where all memory of the extra fifth dimension has been lost. The layer phase actually emerges from  different behaviours in the longitudinal and transverse directions. In fact, it is the result of the strong force in the fifth dimension and the Coulomb force in 4D.
This can be compared to the
 clockwork mechanism where light particles with exponentially suppressed interactions are generated in theories with no fundamental  small parameters.
Both the continuum clockwork and its lattice version we studied here agree
and further supported by the KK spectrum we have calculated.

\vspace{1cm}


\end{document}